\documentclass[preprint,12pt,prb,amsmath,amssymb,floatfix]{revtex4}

\usepackage{bbold}
\usepackage{bm}
\usepackage{color}
\usepackage{dcolumn}
\usepackage{feynmf} 
\usepackage{graphics}
\usepackage{graphicx}
\usepackage{subfigure}
\usepackage{slashed}
\usepackage{placeins}

\def\Tr{\mbox{Tr}}

\def\al{\alpha}

\def\veps{\varepsilon}

\def\be{\begin{equation}}
\def\ee{\end{equation}}
\def\bea{\begin{eqnarray}}
\def\eea{\end{eqnarray}}
\def\bse{\begin{subequations}}
\def\ese{\end{subequations}}
\def\bc{\begin{center}}
\def\ec{\end{center}}

\def\ra{\rightarrow}

\def\nonum{\nonumber}

\def\I{{\rm i}}

\def\D{{\rm d}}
\def\Ord{{\rm O}}

\def\Sp{{\slashed p}}

\def\Sk{{\slashed k}}

\newcommand{\comment}[1]{}

\catcode`,\active

\catcode`\,12

\begin{document}

\title{The method of uniqueness and the optical conductivity of graphene: new application of a powerful technique for multi-loop calculations}


\author{ S.~Teber$^{1,2}$ and A.\ V.~Kotikov$^3$}
\affiliation{
$^1$Sorbonne Universit\'es, UPMC Univ Paris 06, UMR 7589, LPTHE, F-75005, Paris, France.\\
$^2$CNRS, UMR 7589, LPTHE, F-75005, Paris, France.\\
$^3$Bogoliubov Laboratory of Theoretical Physics, Joint Institute for Nuclear Research, 141980 Dubna, Russia.}

\date{\today}

\begin{abstract}
We review the method of uniqueness which is a powerful technique for multi-loop calculations
in higher dimensional theories with conformal symmetry.
We use the method in momentum space and show that it allows a very transparent evaluation
of a two-loop massless propagator Feynman diagram with a non-integer index on the central line.
The result is applied to the computation of the optical conductivity of graphene at the
infra-red Lorentz invariant fixed point. The effect of counter-terms is analysed. A brief comparison with 
the non-relativistic case is included.
\end{abstract}

\maketitle

\begin{fmffile}{fmfmqft-15}

\section{Introduction}

At the heart of perturbative quantum field theory is the exact computation of multi-loop Feynman diagrams.
The later are of crucial importance for the evaluation of renormalization group functions, 
{\it i.e.}, $\beta$-functions and anomalous dimensions of fields, with wide applications ranging from particle physics,
to statistical mechanics and condensed matter physics. At the end of this paper we shall focus on an application 
concerning the optical conductivity of graphene.

From the quantum field theory point of view, since the 1980's, a number of powerful covariant methods have been developed to achieve this task
for dimensionally regularized Feynman diagrams. One of the most widely used methods is integration by parts (IBP) 
which has been introduced by Vasil'ev, Pis'mak and Khonkonen~\cite{VasilievPK81}
and Chetyrkin and Tkachov.~\cite{ChetyrkinT81} It allows to reduce a complicated Feynman diagram in terms of a limited number of so-called ``master integrals'';
such reduction is now automated via it's implementation in computer programs with the help of Laporta's algorithm.~\cite{Laporta00}
In some simple cases, the master integrals themselves can be computed from IBP alone. In general, however, other methods have to be used often in combination with IBP.
A well known method is the Gegenbauer polynomial technique.~\cite{ChetyrkinKT80,Kotikov96} Another very powerful but less popular method is
the method of uniqueness and we shall focus on this method in the following. 
This method owes its name to the so-called uniqueness relation, otherwise known as the star-triangle or Yang-Baxter
relation, which is used in theories with conformal symmetry. Historically,
such relation was probably first used to compute three-dimensional integrals by D'Eramo, Peleti and Parisi.~\cite{DeramoPP71}
Within the framework of multi-loop calculations, the method has first been introduced by Vasil'ev, Pis'mak and Khonkonen.~\cite{VasilievPK81}
It allows, in principle, the computation of complicated Feynman diagrams using sequences of
simple transformations (including integration by parts) without performing any explicit integration.
A diagram is straightforwardly integrated once the appropriate sequence is found. In a sense, the method greatly simplifies
multi-loop calculations.~\cite{VasilievPK81,Usyukina83,Kazakov84,Kazakov85} 
As a matter of fact, the first analytical expression for the
five-loop $\beta$-function of the $\varphi^4$-model was derived by Kazakov using this technique.~\cite{Kazakov84,Kazakov85}
For a given diagram, the task of finding the sequence of transformations is, however, highly non-trivial.

In the following, we will present the method of uniqueness in momentum space in very close analogy with the review of Ref.~[\onlinecite{Kazakov-lectures-84}]
where the method was presented in coordinate space. 
We will then show that it allows a very transparent evaluation
of a two-loop massless propagator Feynman diagram with a non-integer index on the central line.
The result will then be applied to the computation of the optical conductivity of graphene at the
infra-red Lorentz invariant fixed point. In the following, except in Section \ref{sec:application} concerning applications, 
we will work in Euclidean space and set $\hbar=c=1$.

\section{The two-loop massless propagator-type diagram}

\begin{figure}
   \includegraphics[width=0.25\textwidth]{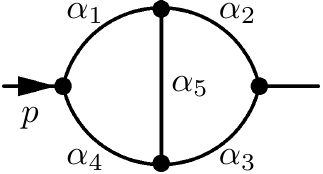}
   \caption{\label{fig:J}
   Two-loop massless propagator diagram.}
\end{figure}
\FloatBarrier

In what follows we shall use dimensional regularization and perform all calculations in an Euclidean space-time of dimensionality
$D=4-2\veps$; alternatively, it will be of practical use to write $D= 2 + 2\lambda$ where $\lambda$ and $\veps$ are related by: $\lambda = 1 - \veps$. 
Our main focus will be on the two-loop massless propagator-type diagram (the so-called p-integral) in momentum space, see Fig.~\ref{fig:J}.
This diagram is one of the building blocks of multi-loop calculations. It reads:
\bea
J(\al_1,\al_2,\al_3,\al_4,\al_5) =
\int \int \frac{\D^D k_1 \, \D^D k_2}{k_1^{2\al_1}\,k_2^{2\al_2}\,(k_2-p)^{2\al_3}\,(k_1-p)^{2\al_4}\,(k_2-k_1)^{2\al_5}} \, ,
\label{def:J}
\eea
with arbitrary indices $\al_i$ and external momentum $p$.
Dimensional analysis gives the dependence of $J$ on $p$ via the dimension $d_F$ of the diagram where: 
\be
d_F = 2(D - \sum_{i=1}^5\,\al_i)\, .
\ee
One of the main goals of multi-loop calculations is to compute the (dimensionless) coefficient function of the diagram that we define as:
\be
I_D(\al_1,\al_2,\al_3,\al_4,\al_5)  \quad = \quad C_D\left[ \quad \parbox{16mm}{
    \begin{fmfgraph*}(16,14)
      \fmfleft{i}
      \fmfright{o}
      \fmfleft{ve}
      \fmfright{vo}
      \fmftop{vn}
      \fmftop{vs}
      \fmffreeze
      \fmfforce{(-0.1w,0.5h)}{i}
      \fmfforce{(1.1w,0.5h)}{o}
      \fmfforce{(0w,0.5h)}{ve}
      \fmfforce{(1.0w,0.5h)}{vo}
      \fmfforce{(.5w,0.95h)}{vn}
      \fmfforce{(.5w,0.05h)}{vs}
      \fmffreeze
      \fmf{plain}{i,ve}
      \fmf{plain,left=0.8}{ve,vo}
      \fmf{phantom,left=0.5,label=$\al_1$,l.d=-0.01w}{ve,vn}
      \fmf{phantom,right=0.5,label=$\al_2$,l.d=-0.01w}{vo,vn}
      \fmf{plain,left=0.8}{vo,ve}
      \fmf{phantom,left=0.5,label=$\al_3$,l.d=-0.01w}{vo,vs}
      \fmf{phantom,right=0.5,label=$\al_4$,l.d=-0.01w}{ve,vs}
      \fmf{plain,label=$\al_5$,l.d=0.05w}{vs,vn}
      \fmf{plain}{vo,o}
      \fmffreeze
      \fmfdot{ve,vn,vo,vs}
    \end{fmfgraph*}
}
\quad \right] \quad = \quad
\frac{p^{d_F}}{\pi^D}\,J(\al_1,\al_2,\al_3,\al_4,\al_5).
\label{def:CF}
\ee
In general, it takes the form of a Laurent series in $\veps$.

The 2-loop massless propagator-type diagram of Eq.~(\ref{def:J}) has a rather long history, see Ref.~[\onlinecite{Grozin}] for a review.
Generally speaking, when all indices are integers the diagram is easily computed, {\it e.g.}, with the help of IBP.
On the other hand, for arbitrary (non-integer) values of all the indices, it's evaluation is highly non-trivial and 
peculiar cases have to be considered, see, {\it e.g.}, 
Refs.~[\onlinecite{ChetyrkinKT80,VasilievPK81,ChetyrkinT81,Kazakov84,Kazakov85,Broadhurst86,BarfootB88,Gracey92,KivelSV93,Kotikov96,BroadhurstGK97,BroadhurstK98,Broadhurst03,BierenbaumW03,KotikovT13,KotikovT14}]. 
In the case where all the indices take the form $\al_i = 1 + a_i \veps$, the diagram is known only in the form of an
$\veps$-expansion; after two decades of calculations,~\cite{ChetyrkinKT80,Kazakov84,Kazakov85,Broadhurst86,BarfootB88,BroadhurstGK97} an expansion to order $\veps^9$ 
was achieved in Ref.~[\onlinecite{Broadhurst03}].
The expansion was automated in Ref.~[\onlinecite{BierenbaumW03}] by representing the diagram as a combination of two-fold series; in principle,
such numerical evaluation allows an expansion to arbitrary order in $\veps$, the only restrictions arising from hardware constraints.
In some other cases, an exact evaluation of the diagram could be found. 
In the case where two adjacent indices are arbitrary (other indices being integers),
the diagram was first computed exactly using the Gegenbauer polynomial technique.~\cite{ChetyrkinKT80}
Such results were soon after recovered in a more simple way using IBP.~\cite{VasilievPK81,ChetyrkinT81}
A decade later, a new class of complicated diagrams where two adjacent indices are integers and the three others are arbitrary
could be computed exactly on the basis of a new development of the Gegenbauer polynomial technique.~\cite{Kotikov96}
For this class of diagrams, similar results have been obtained in Ref.~[\onlinecite{BroadhurstGK97}] using an Ansatz to solve the
recurrence relations arising from IBP for the 2-loop diagram.
All these results are expressed in terms of (combinations of) generalized hypergeometric functions, ${}_3F_2$ with argument $1$.~\cite{Kotikov96,BroadhurstGK97,KotikovT14}
Actually, the simplest non-trivial diagram belonging to this class is the one with an arbitrary index on the central line.
It's coefficient function reads:
\be
I(\al) \quad = \quad 
C_D\left[ \quad \parbox{16mm}{
    \begin{fmfgraph*}(16,14)
      \fmfleft{i}
      \fmfright{o}
      \fmfleft{ve}
      \fmfright{vo}
      \fmftop{vn}
      \fmftop{vs}
      \fmffreeze
      \fmfforce{(-0.1w,0.5h)}{i}
      \fmfforce{(1.1w,0.5h)}{o}
      \fmfforce{(0w,0.5h)}{ve}
      \fmfforce{(1.0w,0.5h)}{vo}
      \fmfforce{(.5w,0.95h)}{vn}
      \fmfforce{(.5w,0.05h)}{vs}
      \fmffreeze
      \fmf{plain}{i,ve}
      \fmf{plain,left=0.8}{ve,vo}
      \fmf{phantom,left=0.5,label=$1$,l.d=-0.01w}{ve,vn}
      \fmf{phantom,right=0.5,label=$1$,l.d=-0.01w}{vo,vn}
      \fmf{plain,left=0.8}{vo,ve}
      \fmf{phantom,left=0.5,label=$1$,l.d=-0.01w}{vo,vs}
      \fmf{phantom,right=0.5,label=$1$,l.d=-0.01w}{ve,vs}
      \fmf{plain,label=$\al$,l.d=0.05w}{vs,vn}
      \fmf{plain}{vo,o}
      \fmffreeze
      \fmfdot{ve,vn,vo,vs}
    \end{fmfgraph*}
}
\quad \right] \quad = \quad
\frac{p^{2(2-\al)}}{\pi^D}\,J(1,1,1,1,\al)\, .
\label{def:I[al]}
\ee
An expression of this diagram in terms of a one-fold series has first been given in Ref.~[\onlinecite{Kazakov85}]. Then, in Ref.~[\onlinecite{Kotikov96}], 
this function was shown to be expressed in terms of a single ${}_3F_2$ function of argument $1$ with the result reading:
\bea
I(\al) &=&
-2\, \Gamma(\lambda)\Gamma(\lambda-\al) \Gamma(1-2\lambda+\al) \times
\label{KotikovG}\\
&\times& \left [ \frac{\Gamma(\lambda)}{\Gamma(2\lambda)\Gamma(3\lambda-\al-1)}\,
\sum_{n=0}^{\infty}\,\frac{\Gamma(n+2\lambda)\Gamma(n+1)}{n!\,\Gamma(n+1+\al)}\,\frac{1}{n+1-\lambda+\al}
+\frac{\pi \cot \pi (2\lambda-\al)}{\Gamma(2\lambda)} \right ]\, .
\nonum
\eea
In these Proceedings, we shall consider an even simpler case where the index $\al$ is related to the dimensionality of the system as follows:
$\al = \lambda = D/2 - 1$. Graphically:
\be
I(\lambda) \quad = \quad
C_D\left[ \quad \parbox{16mm}{
    \begin{fmfgraph*}(16,14)
      \fmfleft{i}
      \fmfright{o}
      \fmfleft{ve}
      \fmfright{vo}
      \fmftop{vn}
      \fmftop{vs}
      \fmffreeze
      \fmfforce{(-0.1w,0.5h)}{i}
      \fmfforce{(1.1w,0.5h)}{o}
      \fmfforce{(0w,0.5h)}{ve}
      \fmfforce{(1.0w,0.5h)}{vo}
      \fmfforce{(.5w,0.95h)}{vn}
      \fmfforce{(.5w,0.05h)}{vs}
      \fmffreeze
      \fmf{plain}{i,ve}
      \fmf{plain,left=0.8}{ve,vo}
      \fmf{phantom,left=0.5,label=$1$,l.d=-0.01w}{ve,vn}
      \fmf{phantom,right=0.5,label=$1$,l.d=-0.01w}{vo,vn}
      \fmf{plain,left=0.8}{vo,ve}
      \fmf{phantom,left=0.5,label=$1$,l.d=-0.01w}{vo,vs}
      \fmf{phantom,right=0.5,label=$1$,l.d=-0.01w}{ve,vs}
      \fmf{plain,label=$\lambda$,l.d=0.05w}{vs,vn}
      \fmf{plain}{vo,o}
      \fmffreeze
      \fmfdot{ve,vn,vo,vs}
    \end{fmfgraph*}
}
\quad \right] \quad = \quad
\frac{p^{2(2-\lambda)}}{\pi^D}\,J(1,1,1,1,\lambda)\, \qquad (\lambda = D/2 - 1)\, .
\label{def:I}
\ee
%
This diagram has already been calculated, see Ref.~[\onlinecite{VasilievPK81}] and also discussions in Ref.~[\onlinecite{KivelSV93}], and reads:
\be
I(\lambda) = 3\,\frac{\Gamma(\lambda)\Gamma(1-\lambda)}{\Gamma(2\lambda)} \,\Big[ \psi'(\lambda) - \psi'(1) \Big] \, .
\label{result:I}
\ee
The method used then was the uniqueness method in position space. In Ref.~[\onlinecite{KotikovT13}], a seemingly simpler and more transparent derivation was proposed 
using the method of uniqueness in momentum space.

\section{The method of uniqueness}  

In what follows all diagrams will be analyzed in momentum space. We assume that algebraic manipulations related to gamma matrices have
been done, {\it e.g.}, contraction of Lorentz indices, calculations of traces, etc... The diagrams we shall consider therefore involve only scalar propagators
which are simple power laws of the form: $1/k^{2\al}$ where $\al$ is the so-called index of the propagators.
The latter can be represented graphically as a line:
\be
\parbox{10mm}{
  \begin{fmfgraph*}(10,10)
    \fmfleft{i}
    \fmfright{o}
    \fmf{plain,label=$\al$,l.s=left}{i,o}
  \end{fmfgraph*}
} \quad \Rightarrow \quad \frac{1}{k^{2\al}}\,.
\label{def:index}
\ee
The index of a diagram is defined as the sum of the indices of its constituent lines. 
Ordinary lines have index $1$, while ordinary triangles and vertices have index $3$.
Of importance in the following will be the notions of unique triangle and unique vertex.
In momentum space a triangle and a vertex are said to be unique if their indices are 
equal to $D=2+2\lambda$ and $D/2=1+\lambda$, respectively, see table \ref{tab:indices} for a summary.

\begin{center}
\renewcommand{\tabcolsep}{0.25cm}
\renewcommand{\arraystretch}{1.5}
\begin{table}
    \begin{tabular}{  l || c | c | c }
      \hline
        ~~      &       {\bf Line}                      &       {\bf Triangle}          &       {\bf Vertex} \\
      \hline \hline
      Arbitrary &       $\al$                           & $\sum_{i=1}^3 \al_i$          &       $\sum_{i=1}^3 \al_i$    \\
      \hline
      Ordinary  &       $1$                             & $3$                           &       $3$     \\
      \hline
      Unique    &       $D/2~=~2-\veps~=~1+\lambda$     & $D~=~4-2\veps~=~2+2\lambda$   &       $D/2 ~=~2-\veps~=~1+\lambda$  \\
      \hline
    \end{tabular}
    \caption{Indices of lines, triangles and vertices in $p$-space ($\veps=2-D/2$, $\lambda = D/2-1$)}
    \label{tab:indices}
\end{table}
\end{center}

With these definitions and graphical notations, chains reduce to the product of propagators:
\be
\parbox{15mm}{
  \begin{fmfgraph*}(15,10)
    \fmfleft{i}
    \fmfright{o}
    \fmf{plain,label=$\al$,l.s=left}{i,v}
    \fmf{plain,label=$\beta$,l.s=left}{v,o}
    \fmfdot{v}
  \end{fmfgraph*}
} \quad = \quad
\parbox{15mm}{
  \begin{fmfgraph*}(15,10)
    \fmfleft{i}
    \fmfright{o}
    \fmf{plain,label=$\al+\beta$,l.s=left}{i,o}
  \end{fmfgraph*}
}\,\quad . 
\label{def:chain}
\ee
On the other hand simple loops involve an integration:
\be
\parbox{15mm}{
    \begin{fmfgraph*}(15,13)
      \fmfleft{i}
      \fmfright{o}
      \fmfleft{ve}
      \fmfright{vo}
      \fmffreeze
      \fmfforce{(-0.3w,0.5h)}{i}
      \fmfforce{(1.3w,0.5h)}{o}
      \fmfforce{(0w,0.5h)}{ve}
      \fmfforce{(1.0w,0.5h)}{vo}
      \fmffreeze
      \fmf{plain}{i,ve}
      \fmf{plain,left=0.6,label=$\al$,l.d=0.1h}{ve,vo}
      \fmf{plain,left=0.6,label=$\beta$,l.d=0.05h}{vo,ve}
      \fmf{plain}{vo,o}
      \fmffreeze
      \fmfdot{ve,vo}
    \end{fmfgraph*}
}
 \qquad = \quad \pi^{D/2} A(\al,\beta)\quad
\parbox{15mm}{
  \begin{fmfgraph*}(15,10)
    \fmfleft{i}
    \fmfright{o}
    \fmf{plain,label=$\al+\beta-D/2$,l.s=left}{i,o}
    \fmfdot{i,o}
  \end{fmfgraph*}
}\quad , 
\label{loop}
\ee
where
\be
A(\al,\beta) = \frac{a(\al)a(\beta)}{a(\al+\beta-D/2)}, \quad a(\al) = \frac{\Gamma(D/2-\al)}{\Gamma(\al)} \, .
\label{def:A}
\ee

Multi-loop diagrams can then be evaluated without further explicit integration by using the above results in combination with 
some identities between diagrams. Central to this paper is the uniqueness (or star-triangle)
relation which relates a so-called unique triangle to a unique vertex:
\bea
\parbox{16mm}{
  \begin{fmfgraph*}(16,16)
    \fmfleft{l}
    \fmfleft{vl}
    \fmfright{r}
    \fmfright{vr}
    \fmftop{t}
    \fmftop{vt}
    \fmffreeze
    \fmfforce{(-0.1w,0.042265h)}{l}
    \fmfforce{(0.1w,0.2h)}{vl}
    \fmfforce{(1.1w,0.042265h)}{r}
    \fmfforce{(0.9w,0.2h)}{vr}
    \fmfforce{(.5w,1.1h)}{t}
    \fmfforce{(.5w,0.966025h)}{vt}
    \fmffreeze
    \fmf{plain,label=$\al_2$,label.side=left,l.d=0.05w}{vr,vl}
    \fmf{plain,label=$\al_3$,label.side=left,l.d=0.05w}{vl,vt}
    \fmf{plain,label=$\al_1$,label.side=left,l.d=0.05w}{vt,vr}
    \fmf{plain}{l,vl}
    \fmf{plain}{r,vr}
    \fmf{plain}{t,vt}
    \fmffreeze
    \fmfdot{vl,vt,vr}
  \end{fmfgraph*}
} \qquad \underset{\underset{i}{\sum} \al_i = D}{=} \qquad \pi^{D/2} A(\al_1,\al_2)\,
\parbox{16mm}{
  \begin{fmfgraph*}(16,16)
    \fmfleft{l}
    \fmfright{r}
    \fmftop{t}
    \fmfleft{v}
    \fmffreeze
    \fmfforce{(-0.1w,0.142265h)}{l}
    \fmfforce{(1.1w,0.142265h)}{r}
    \fmfforce{(.5w,1.1h)}{t}
    \fmfforce{(.5w,0.488675h)}{v}
    \fmffreeze
    \fmf{plain,label=$\tilde{\al}_2$,label.side=left,l.d=0.05w}{v,t}
    \fmf{plain,label=$\tilde{\al}_3$,label.side=right,l.d=0.05w}{r,v}
    \fmf{plain,label=$\tilde{\al}_1$,label.side=left,l.d=0.05w}{v,l}
    \fmffreeze
    \fmfdot{v}
  \end{fmfgraph*}
}\, \qquad ,
\label{def:star-triangle}
\eea
where $\tilde{\al}_i=D/2-\al_i$ is the index dual to $\al_i$. Another important identity takes the form of
a recurrence relation and can be obtained from integration by parts~\cite{VasilievPK81,ChetyrkinT81}: 
\bea
(D-\al_2-\al_3-2\al_5) \quad
\parbox{16mm}{
    \begin{fmfgraph*}(16,14)
      \fmfleft{i}
      \fmfright{o}
      \fmfleft{ve}
      \fmfright{vo}
      \fmftop{vn}
      \fmftop{vs}
      \fmffreeze
      \fmfforce{(-0.1w,0.5h)}{i}
      \fmfforce{(1.1w,0.5h)}{o}
      \fmfforce{(0w,0.5h)}{ve}
      \fmfforce{(1.0w,0.5h)}{vo}
      \fmfforce{(.5w,0.95h)}{vn}
      \fmfforce{(.5w,0.05h)}{vs}
      \fmffreeze
      \fmf{plain}{i,ve}
      \fmf{plain,left=0.8}{ve,vo}
      \fmf{phantom,left=0.5,label=$\al_1$,l.d=-0.01w}{ve,vn}
      \fmf{phantom,right=0.5,label=$\al_2$,l.d=-0.01w}{vo,vn}
      \fmf{plain,left=0.8}{vo,ve}
      \fmf{phantom,left=0.5,label=$\al_3$,l.d=-0.01w}{vo,vs}
      \fmf{phantom,right=0.5,label=$\al_4$,l.d=-0.01w}{ve,vs}
      \fmf{plain,label=$\al_5$,l.d=0.05w}{vs,vn}
      \fmf{plain}{vo,o}
      \fmffreeze
      \fmfdot{ve,vn,vo,vs}
    \end{fmfgraph*}
}
\quad & = & \,\, \al_2 \left[\quad
\parbox{16mm}{
    \begin{fmfgraph*}(16,14)
      \fmfleft{i}
      \fmfright{o}
      \fmfleft{ve}
      \fmfright{vo} 
      \fmftop{vn}
      \fmftop{vs}
      \fmffreeze
      \fmfforce{(-0.1w,0.5h)}{i}
      \fmfforce{(1.1w,0.5h)}{o}
      \fmfforce{(0w,0.5h)}{ve}
      \fmfforce{(1.0w,0.5h)}{vo}
      \fmfforce{(.5w,0.95h)}{vn}
      \fmfforce{(.5w,0.05h)}{vs}
      \fmffreeze
      \fmf{plain}{i,ve}
      \fmf{plain,left=0.8}{ve,vo}
      \fmf{phantom,left=0.5}{ve,vn}
      \fmf{phantom,right=0.4,label=$+$,l.d=-0.01w}{vo,vn}
      \fmf{plain,left=0.8}{vo,ve}
      \fmf{phantom,left=0.5}{vo,vs}
      \fmf{phantom,right=0.5}{ve,vs}
      \fmf{plain,label=$-$,l.d=0.05w}{vs,vn}
      \fmf{plain}{vo,o}
      \fmffreeze
      \fmfdot{ve,vn,vo,vs}
    \end{fmfgraph*}
} \qquad - \qquad
\parbox{16mm}{
    \begin{fmfgraph*}(16,14)
      \fmfleft{i}
      \fmfright{o}
      \fmfleft{ve}
      \fmfright{vo} 
      \fmftop{vn}
      \fmftop{vs}
      \fmffreeze
      \fmfforce{(-0.1w,0.5h)}{i}
      \fmfforce{(1.1w,0.5h)}{o}
      \fmfforce{(0w,0.5h)}{ve}
      \fmfforce{(1.0w,0.5h)}{vo}
      \fmfforce{(.5w,0.95h)}{vn}
      \fmfforce{(.5w,0.05h)}{vs}
      \fmffreeze
      \fmf{plain}{i,ve}
      \fmf{plain,left=0.8}{ve,vo}
      \fmf{phantom,left=0.4,label=$-$,l.d=-0.01w}{ve,vn}
      \fmf{phantom,right=0.4,label=$+$,l.d=-0.01w}{vo,vn}
      \fmf{plain,left=0.8}{vo,ve}
      \fmf{phantom,left=0.5}{vo,vs}
      \fmf{phantom,right=0.5}{ve,vs}
      \fmf{plain}{vs,vn}
      \fmf{plain}{vo,o}
      \fmffreeze
      \fmfdot{ve,vn,vo,vs}
    \end{fmfgraph*}
} \quad
\right]
\nonum
\\ &\qquad& \nonum 
\\ & + & \,\,
\al_3 \left[\quad
\parbox{16mm}{
    \begin{fmfgraph*}(16,14)
      \fmfleft{i}
      \fmfright{o}
      \fmfleft{ve}
      \fmfright{vo}
      \fmftop{vn}
      \fmftop{vs}
      \fmffreeze
      \fmfforce{(-0.1w,0.5h)}{i}
      \fmfforce{(1.1w,0.5h)}{o}
      \fmfforce{(0w,0.5h)}{ve}
      \fmfforce{(1.0w,0.5h)}{vo}
      \fmfforce{(.5w,0.95h)}{vn}
      \fmfforce{(.5w,0.05h)}{vs}
      \fmffreeze
      \fmf{plain}{i,ve}
      \fmf{plain,left=0.8}{ve,vo}
      \fmf{phantom,left=0.5}{ve,vn}
      \fmf{phantom,right=0.5}{vo,vn}
      \fmf{plain,left=0.8}{vo,ve}
      \fmf{phantom,left=0.4,label=$+$,l.d=-0.01w}{vo,vs}
      \fmf{phantom,right=0.5}{ve,vs}
      \fmf{plain,label=$-$,l.d=0.05w}{vs,vn}
      \fmf{plain}{vo,o}
      \fmffreeze
      \fmfdot{ve,vn,vo,vs}
    \end{fmfgraph*}
} \qquad - \qquad
\parbox{16mm}{
    \begin{fmfgraph*}(16,14)
      \fmfleft{i}
      \fmfright{o}
      \fmfleft{ve}
      \fmfright{vo}
      \fmftop{vn}
      \fmftop{vs}
      \fmffreeze
      \fmfforce{(-0.1w,0.5h)}{i}
      \fmfforce{(1.1w,0.5h)}{o}
      \fmfforce{(0w,0.5h)}{ve}
      \fmfforce{(1.0w,0.5h)}{vo}
      \fmfforce{(.5w,0.95h)}{vn}
      \fmfforce{(.5w,0.05h)}{vs}
      \fmffreeze
      \fmf{plain}{i,ve}
      \fmf{plain,left=0.8}{ve,vo}
      \fmf{phantom,left=0.5}{ve,vn}
      \fmf{phantom,right=0.5}{vo,vn}
      \fmf{plain,left=0.8}{vo,ve}
      \fmf{phantom,left=0.4,label=$+$,l.d=-0.01w}{vo,vs}
      \fmf{phantom,right=0.4,label=$-$,l.d=-0.01w}{ve,vs}
      \fmf{plain}{vs,vn}
      \fmf{plain}{vs,vn}
      \fmf{plain}{vo,o}
      \fmffreeze
      \fmfdot{ve,vn,vo,vs}
    \end{fmfgraph*}
} \quad
\right]\, ,
\label{def:IBP}
\eea
where $\pm$ on the right-hand side of the equation denotes the increase or decrease of a line index by $1$ with respect to its value on the left-hand side.
In the following, in order to simply notations, we will assume that lines with no index are ordinary lines, that is lines of index $1$.

\section{Calculation of the diagram} 

With the help of the above notations and identities, $I(\lambda)$ can be computed. 
We only give the elementary four steps in what follows, see Ref.~[\onlinecite{KotikovT13}] for more details.
The first step consists in replacing the central line by a 
loop, Eq.~(\ref{loop}), in order to make the right triangle unique. 
Then, the uniqueness relation, Eq.~(\ref{def:star-triangle}), can then be used. 
In graphical notations these two steps read:
\be
J(1,1,1,1,\lambda) \,\, = \quad
\parbox{18mm}{
  \begin{fmfgraph*}(18,16)
    \fmfleft{i}
    \fmfright{o}
    \fmfleft{ve}
    \fmfright{vo}
    \fmftop{vn}
    \fmftop{vs}    
    \fmffreeze
    \fmfforce{(-0.1w,0.5h)}{i}
    \fmfforce{(1.1w,0.5h)}{o}
    \fmfforce{(0w,0.5h)}{ve}
    \fmfforce{(1.0w,0.5h)}{vo}
    \fmfforce{(.5w,0.95h)}{vn}
    \fmfforce{(.5w,0.05h)}{vs}
    \fmffreeze
    \fmf{plain}{i,ve}
    \fmf{plain,left=0.8}{ve,vo}
    \fmf{plain,left=0.8}{vo,ve}
    \fmf{plain,label=$\lambda$,l.d=0.05w}{vs,vn}
    \fmf{plain}{vo,o}
    \fmffreeze
    \fmfdot{ve,vn,vo,vs}
  \end{fmfgraph*}
} \quad  =  \, \frac{1}{\pi^{D/2} A(1,2\lambda)} \quad
\parbox{18mm}{
  \begin{fmfgraph*}(18,16)
    \fmfleft{i}
    \fmfright{o}
    \fmfleft{ve}
    \fmfright{vo}
    \fmftop{vn}
    \fmftop{vs}
    \fmffreeze
    \fmfforce{(-0.1w,0.5h)}{i}
    \fmfforce{(1.1w,0.5h)}{o}
    \fmfforce{(0w,0.5h)}{ve}
    \fmfforce{(1.0w,0.5h)}{vo}
    \fmfforce{(.5w,0.95h)}{vn}
    \fmfforce{(.5w,0.05h)}{vs}
    \fmffreeze
    \fmf{plain}{i,ve}
    \fmf{plain,left=0.8}{ve,vo}
    \fmf{plain,left=0.8}{vo,ve}
    \fmf{plain,left=0.3}{vs,vn}
    \fmf{plain,left=0.3,label=$2\lambda$,l.d=0.05w}{vn,vs}
    \fmf{plain}{vo,o}
    \fmffreeze
    \fmfdot{ve,vn,vo,vs}
  \end{fmfgraph*}
} \quad = \quad
\parbox{16mm}{
  \begin{fmfgraph*}(16,16)
    \fmfleft{i}
    \fmfright{o}
    \fmfleft{ve}
    \fmfright{vo}
    \fmftop{vn}
    \fmftop{vs}
    \fmffreeze
    \fmfforce{(-0.1w,0.5h)}{i}
    \fmfforce{(1.1w,0.5h)}{o}
    \fmfforce{(0w,0.5h)}{ve}
    \fmfforce{(1.0w,0.5h)}{vo}
    \fmfforce{(.5w,0.9h)}{vn}
    \fmfforce{(.5w,0.1h)}{vs}
    \fmffreeze
    \fmf{plain}{i,ve}
    \fmf{plain,left=0.8}{ve,vo}
    \fmf{plain,left=0.8}{vo,ve}
    \fmf{plain}{vs,vn}
    \fmf{phantom,right,label=$\lambda$,l.s=left,l.d=-0.05h}{vo,vn}
    \fmf{phantom,left,label=$\lambda$,l.s=right,l.d=-0.05h}{vo,vs}
    \fmf{plain}{vo,o}
    \fmffreeze
    \fmfdot{ve,vn,vo,vs}
  \end{fmfgraph*}
} \quad \frac{1}{p^{2(1-\lambda)}} \, .
\label{first+second-steps}
\ee
The third step consists in using the integration by parts identity, Eq.~(\ref{def:IBP}),  in order to reduce the last diagram to sequences of chains and 
simple loops: 
\bea
( -2 \delta ) \quad
\parbox{16mm}{
  \begin{fmfgraph*}(16,16)
    \fmfleft{i}
    \fmfright{o}
    \fmfleft{ve}
    \fmfright{vo}
    \fmftop{vn}
    \fmftop{vs}
    \fmffreeze
    \fmfforce{(-0.1w,0.5h)}{i}
    \fmfforce{(1.1w,0.5h)}{o}
    \fmfforce{(0w,0.5h)}{ve}
    \fmfforce{(1.0w,0.5h)}{vo}
    \fmfforce{(.5w,0.9h)}{vn}
    \fmfforce{(.5w,0.1h)}{vs}
    \fmffreeze
    \fmf{plain}{i,ve}
    \fmf{plain,left=0.8}{ve,vo}
    \fmf{plain,left=0.8}{vo,ve}
    \fmf{plain}{vs,vn}
    \fmf{phantom,right=0.1,label=$\lambda+\delta$,l.s=right}{vo,vn}
    \fmf{phantom,left=0.1,label=$\lambda+\delta$,l.s=left}{vo,vs}
    \fmf{plain}{vo,o}
    \fmffreeze
    \fmfdot{ve,vn,vo,vs}
  \end{fmfgraph*}
} \, \, &=& \,\, 2 ( \lambda + \delta ) \quad 
\left[ \quad
\parbox{32mm}{
  \begin{fmfgraph*}(32,16)
    \fmfleft{i}
    \fmfright{o}
    \fmfleft{ve}
    \fmfright{vo}
    \fmftop{v}
    \fmffreeze
    \fmfforce{(-0.1w,0.5h)}{i}
    \fmfforce{(1.1w,0.5h)}{o}
    \fmfforce{(0w,0.5h)}{ve}
    \fmfforce{(1.0w,0.5h)}{vo}
    \fmfforce{(.5w,0.5h)}{v}
    \fmffreeze
    \fmf{plain}{i,ve}
    \fmf{plain,left=0.8}{ve,v}
    \fmf{plain,left=0.8}{v,ve}
    \fmf{plain,left=0.8,label=$\lambda+\delta$,l.s=left}{v,vo}
    \fmf{plain,left=0.8,label=$\lambda+\delta+1$,l.s=left}{vo,v}
    \fmf{plain}{vo,o}
    \fmffreeze
    \fmfdot{ve,v,vo}
  \end{fmfgraph*}
} \qquad - \qquad
\parbox{16mm}{
  \begin{fmfgraph*}(16,16)
    \fmfleft{i}
    \fmfright{o}
    \fmfleft{ve}
    \fmfright{vo}
    \fmftop{v}
    \fmffreeze
    \fmfforce{(-0.1w,0.5h)}{i}
    \fmfforce{(1.1w,0.5h)}{o}
    \fmfforce{(0w,0.5h)}{ve}
    \fmfforce{(1.0w,0.5h)}{vo}
    \fmfforce{(.5w,0.9h)}{v}
    \fmffreeze
    \fmf{plain}{i,ve}
    \fmf{plain,left=0.8}{ve,vo}
    \fmf{plain,left=0.8,label=$\lambda+\delta+1$,l.s=left}{vo,ve}
    \fmf{plain,left=0.5}{v,ve}
    \fmf{phantom,right=0.1,label=$\lambda+\delta$,l.s=right}{vo,v}
    \fmf{plain}{vo,o}
    \fmffreeze
    \fmfdot{ve,v,vo}
  \end{fmfgraph*}
} \qquad
\right] \,\, ,
\label{third-step} \\
&\quad&
\nonumber 
\eea
The resulting diagrams can then be immediately integrated with the help of Eqs.~(\ref{def:chain}) and
(\ref{loop}). Notice that in Eq.~(\ref{third-step}) an additional regularization parameter $\delta$ has been introduced in order to make sense of the 
integration by parts identity. The fourth and last step then consists in taking the limit $\delta \ra 0$ which can be done by using the following product expansion of the Gamma function:
\be
\Gamma(x+\veps) = \Gamma(x)\,\exp \Big[ \,\,\sum_{k=1}^{\infty} \psi^{(k-1)}(x) \frac{\veps^k}{k!} \,\,\Big]\, ,
\quad
\psi(x) = \psi^{(0)}(x) = \frac{\Gamma'(x)}{\Gamma(x)}, \quad  \psi^{(k)}(x) = \frac{\D^k}{\D x^k} \,\psi(x)\, ,
\label{Gamma-expansion}
\ee
where $\psi^{(k)}$ is the polygamma function of order $k$. All calculations done, this yields:
\be
\parbox{16mm}{
  \begin{fmfgraph*}(16,16)
    \fmfleft{i}
    \fmfright{o}
    \fmfleft{ve}
    \fmfright{vo}
    \fmftop{vn}
    \fmftop{vs}
    \fmffreeze
    \fmfforce{(-0.1w,0.5h)}{i}
    \fmfforce{(1.1w,0.5h)}{o}
    \fmfforce{(0w,0.5h)}{ve}
    \fmfforce{(1.0w,0.5h)}{vo}
    \fmfforce{(.5w,0.9h)}{vn}
    \fmfforce{(.5w,0.1h)}{vs}
    \fmffreeze
    \fmf{plain}{i,ve}
    \fmf{plain,left=0.8}{ve,vo}
    \fmf{plain,left=0.8}{vo,ve}
    \fmf{plain}{vs,vn}
    \fmf{phantom,right=0.1,label=$\lambda$,l.s=right}{vo,vn}
    \fmf{phantom,left=0.1,label=$\lambda$,l.s=left}{vo,vs}
    \fmf{plain}{vo,o}
    \fmffreeze
    \fmfdot{ve,vn,vo,vs}
  \end{fmfgraph*}
} \quad = \, \frac{\pi^D}{p^2}\,3\,\frac{\Gamma(\lambda)\Gamma(1-\lambda)}{\Gamma(2\lambda)} \,\Big[ \psi'(\lambda) - \psi'(1) \Big] \, ,
\label{inter}
\ee
where $\psi'(x)$ is the trigamma function.
Substituting this result in Eq.~(\ref{first+second-steps}) and using Eq.~(\ref{def:I}), we obtain the advertised result\cite{VasilievPK81,KivelSV93} for the coefficient function, Eq.~(\ref{result:I}). 
In the even-dimensional case ($\lambda \ra 1$ or $D \ra 4$) the well-known result: 
\be
I(1) = 6 \, \zeta(3)\, ,
\ee
is recovered. On the other hand, in the odd-dimensional case ($\lambda \ra 1/2$ or $D \ra 3$), which is one of the cases of interest to Ref.~[\onlinecite{KotikovT13,Teber12,KivelSV93}], 
the result is transcendentally more complex ($\zeta(2) = \pi^2/6$) and reads: 
\be
I(1/2) = 6 \pi \,\zeta(2)\, .
\ee

\section{Application} 
\label{sec:application}


Going back to Minkowski space, we consider the following low-energy effective action describing 
the coupling of a fermion field in $d_e=D_e+1$-dimensions with a $U(1)$ gauge field in $d_\gamma = D_\gamma + 1$-dimensions:
\bea
S =&& \sum_{\sigma=1}^{N_F} \int \D t\, \D^{D_e} x\, \left[ \bar{\psi}_\sigma \left( \I \gamma^0 \partial_t + \I v \vec{\gamma} \cdot \vec{\nabla}\,\right) \psi_\sigma - e\bar{\psi}_\sigma \,\gamma^0 A_0\, \psi_\sigma
+ e \frac{v}{c}\,\bar{\psi}_\sigma\, \vec{\gamma} \cdot \vec{A}\, \psi_\sigma \right ] 
\nonum \\
&&+\, \int \D t\, \D^{D_\gamma} x\,\left[ - \frac{1}{4}\,F^{\mu \nu}\,F_{\mu \nu} - \frac{1}{2a}\left(\partial_{\mu}A^{\mu}\right)^2 \right]\, ,
\label{model}
\eea
where $\psi_\sigma$ is a four component spinor of spin index $\sigma$ which varies from $1$ to $N_F$ ($N_F=2$ for graphene), 
$v$ is the Fermi velocity, $c$ is the velocity of light which is also implicitly contained in the gauge field action through $\partial_\mu = (\frac{1}{c}\partial_t,\vec \nabla\,)$, 
$a$ is the gauge fixing parameter and $\gamma^\mu$ is a $4\times 4$ Dirac matrix satisfying the
usual algebra: $\{ \gamma^\mu,\gamma^\nu \} = 2 g^{\mu \nu}$
where $g^{\mu \nu} = {\rm diag}(1,-1,-1,\cdots,-1)$ is the metric tensor in $D_e+1$-dimensions.
As we shall use dimensional regularization, the following parametrisation will be useful:~\footnote{The parameter $\lambda$ here is related to the parameter $\veps_e$ of Ref.~[\onlinecite{Teber12}] via the relation: $\lambda = 1 -\veps_e$.}
\be
d_\gamma = 4 -2\veps_\gamma, \qquad d_e = 2 + 2 \lambda - 2\veps_\gamma\, .
\ee
The case of graphene corresponds to: $\veps_\gamma \ra 0$ and $\lambda \ra 1/2$, that is a fermion living in a space of $d_e = 2 +1$-dimensions
interacting with a gauge field in $d_\gamma=3+1$-dimensions.

From the action of Eq.~(\ref{model}) and for arbitrary $d_\gamma$ and $d_e$, we have the following Feynman rules. The fermion propagator reads:
\be
S_0(p) = \frac{\I \Sp}{p^2}\, , \qquad \Sp = \gamma^\mu p_\mu = \gamma^0 p_0 - v \vec{\gamma}\cdot \vec{p}\, .
\label{fermion-prop0}
\ee
The reduced photon propagator is given by:
\be
\tilde{D}_0^{\mu \nu}(\bar{q}) =
\frac{\I}{(4\pi)^{1-\lambda}}\frac{\Gamma(\lambda)}{(-\bar{q}^{\,2})^{\lambda}}\,\left( g^{\mu \nu} - \tilde{\xi}\,\frac{\bar{q}^{\,\mu} \, \bar{q}^{\,\nu}}{\bar{q}^{\,2}} \right)\, ,
\qquad \bar{q}^{\,\mu} = ( q^0/c , \vec{q}\,)\, ,
\label{gauge-field-prop0}
\ee
where $\tilde{\xi} = \lambda\,\xi = \lambda\,(1-a)$. It is a reduced propagator since it has been obtained from the usual $D_\gamma$-dimensional photon propagator
after integrating out all space coordinates perpendicular to the $D_e$-dimensional membrane.
Finally, the vertex function ($j=1,2$ is a space index) reads:
\be
\Gamma_0^0 = -\I e \gamma^0 \, , \qquad \Gamma_0^j = -\I e \frac{v}{c} \gamma^j\, .
\ee
The coupling of the fermion field to the gauge field is characterized by a dimensionless coupling constant: $\al_g = e^2/(4\pi v)$. 
In graphene, a well known effect of interactions is to renormalize the Fermi velocity which then flows to the velocity of light in the infra-red, see Ref.~[\onlinecite{GonzalezGV93}].
Experimentally,\cite{Elias11} the flow of the velocity is cut at values of the order of $v \approx c/300$.
Hence, it is common practice to neglect all retardation effects. In this non-relativistic limit there is no coupling of the fermion to the vector photon ($\Gamma_0^j=0$) and the rules (for arbitrary $\lambda$) simplify as:
\be
S_0(p) = \frac{\I \Sp}{p^2}, \quad \tilde{D}_0^{0 0}(\vec q\,) = \frac{\I}{(4\pi)^{1-\lambda}}\,\frac{\Gamma(\lambda)}{(|\vec q\,|^2)^\lambda}, \quad \Gamma_0^0 = -\I e \gamma^0 \qquad (v \ll c)\, ,
\ee
where $p^\mu = (p^0,v\vec p\,)$. In this limit, the coupling constant is of the order of unity: $\al_g \approx 300/137 \approx 2$.
Formally, we may let the velocity flow. Then, deep in the infra-red, a Lorentz invariant fixed point is reached:\cite{GonzalezGV93} 
$v \ra c$ and $\al_g \ra \al \approx 1/137$.
At the fixed point, the rules (again for arbitrary $\lambda$) simplify and read:
\be
S_0(p) = \frac{\I \Sp}{p^2}, \quad \tilde{D}_0^{\mu \nu}(q) =
\frac{\I}{(4\pi)^{1-\lambda}}\,\frac{\Gamma(\lambda)}{(-q^{2})^{1/2}}\,\left( g^{\mu \nu} - \tilde{\xi}\,\frac{q^{\,\mu} \,q^{\nu}}{q^{\,2}} \right), \quad \Gamma^\mu = -\I e \gamma^\mu \qquad (v=c)\, ,
\ee
where momenta are of the form $k^\mu = (k^0,\vec k\,)$ as we may set $v=c=1$ in this limit because there is no further flow of the velocity.
This relativistic model belongs to the class of the so-called massless reduced quantum electrodynamics (reduced QED), [\onlinecite{GorbarGM01}], or
pseudo QED, Refs.~[\onlinecite{Marino93+DoreyM92+KovnerR90}], models. The Lorentz invariance allows a straightforward application 
of covariant methods for computing massless Feynman diagrams. The square root branch cut in the photon propagator leads to non trivial Feynman diagrams with
non-integer indices of the type considered in the previous sections. 


Given the above rules, we now focus on the computation of radiative corrections to the polarization operator 
$\Pi^{\mu\nu}(q)$. Power counting indicates that this function is free from singularities in the limit $\veps_\gamma \ra 0$ and $\lambda \ra 1/2$ that we are interested in.
In the following, we shall therefore perform all calculations for $\veps_\gamma \ra 0$ keeping arbitrary $\lambda$. At the end, the case $\lambda=1/2$ will
be considered. Similar computations valid for any $\veps_\gamma$ and $\lambda$ can be found in Ref.~[\onlinecite{Teber12}].
Once computed, the polarization operator can be related to an observable, the so-called optical conductivity, with the help of the following relation:
\be
\sigma(q_0) = - \lim_{\vec{q} \ra 0} \, \frac{\I q_0}{|\vec{q}\,|^2}\,\Pi^{00}(q_0,\vec{q}\,)\, .
\label{sigma00}
\ee
Equivalently, from the parametrization $\Pi^{\mu\nu}(q) = \Pi(q^2) \, (g^{\mu\nu} q^2 - q^{\mu}q^{\nu})$ which follows from current conservation, 
we may write $\sigma(q_0) = \I q_0\, v^2\,\Pi(q_0^2, \vec{0}\,)$. In the non-relativistic limit, the computation of $\Pi^{00}(q)$ is the most convenient
while in the relativistic limit it is that of $\Pi(q^2)$. 

\begin{figure}
  \includegraphics{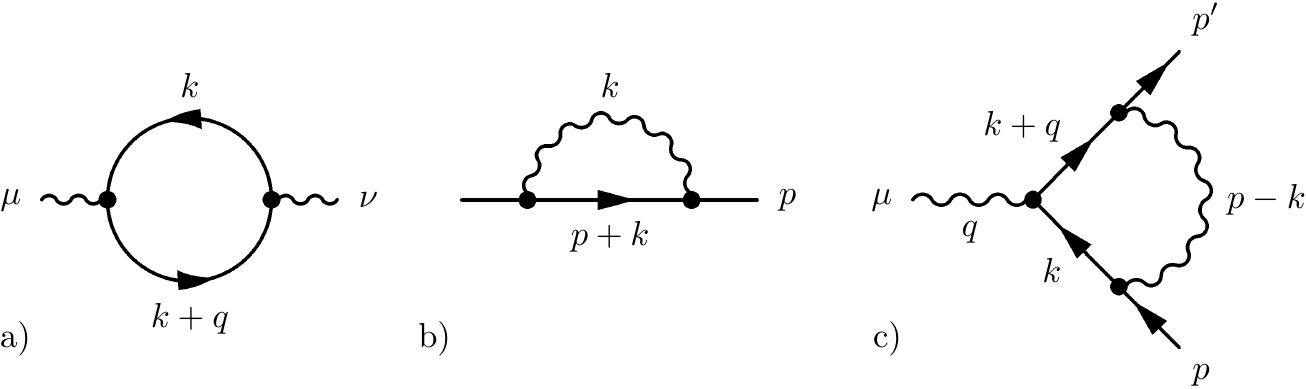}
  \caption{\label{fig:one-loop}
  One-loop diagrams: a) vacuum polarization, b) fermion self-energy and c) fermion-gauge field vertex.}
\end{figure}

In the absence of electron-electron interactions, the polarization operator reduces to a simple loop, see Fig.~\ref{fig:one-loop}a, which is defined as:
\be
i\Pi_1^{\mu \nu}(q) = - \int [\D^{d_e}k]\, \Tr \left[ (-ie\gamma^\mu) S_0(k+q)(-ie\gamma^\nu)S_0(k) \right]\, , \qquad [\D^{d_e}k] = \frac{\D^{d_e}k}{(2\pi)^{d_e}}\, .
\label{pi-1l}
\ee
Focusing on the computation of $\Pi_{1}(q^2)$, straightforward calculations lead to (for $\veps_\gamma \ra 0$):
\be
\Pi_{1}(q^2) = - 4N_F\,\frac{e^2 \, \Gamma(\lambda)}{(4\pi)^{1+\lambda} \, \left(-q^2\right)^{1-\lambda}}
\frac{\Gamma(1+\lambda) \Gamma(1-\lambda)}{ (1+2\lambda) \Gamma(2\lambda)} \, .
\label{1loop}
\ee
In the case $\lambda=1/2$ which is of interest to us, this yields: $\Pi_1(q^2) = - e^2 N_F/(8\, \sqrt{-q^2})$. Hence, we
recover the minimal conductivity of graphene ($N_F=2$):~\cite{LudwigFSG94} 
\be
\sigma_0 = e^2/(4 \hbar)\, .
\label{sigma0}
\ee
Because this is a free fermion result, it holds for any value of $v/c$. As we shall now see, this is not so for higher order corrections.

\begin{figure}
    \includegraphics[width=0.75\textwidth]{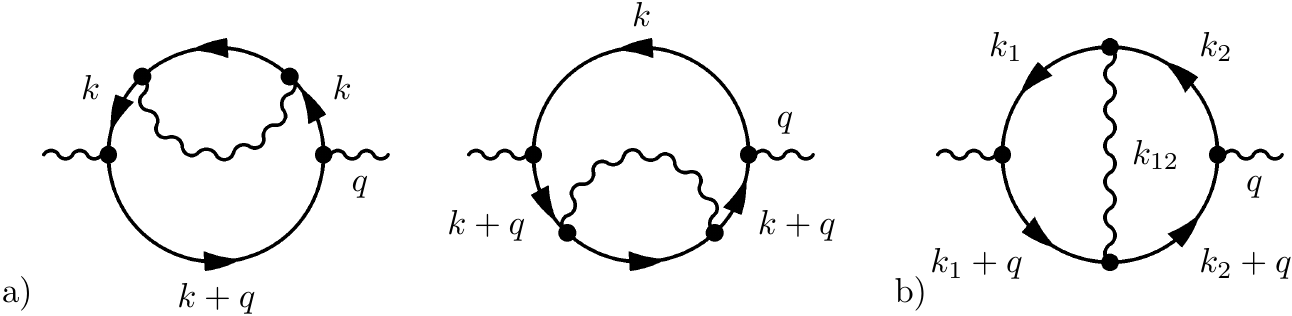}
    \caption{\label{fig:rqed-2loop-polarization}
     Two-loop vacuum polarization diagrams.}
\end{figure}

In the last decade, there has been extensive work done in order to understand how electron-electron interactions
may affect the result of Eq.~(\ref{sigma0}), see Refs.~[\onlinecite{Mishchenko08,C-theory,TK14,GiulianiMP11,HerbutM13}].  
Experiments~\cite{C-experiments} seem to indicate that interaction effects are weak and lead only to $2\%$ 
deviation from the free fermion result, see also Ref.~[\onlinecite{Peres10}] for a review.
Theoretically, it was shown that there are no corrections to the optical conductivity from short-range interactions among the fermions.\cite{GiulianiMP11}
No exact result is available in the case of long-range interactions that we are interested in. In this case, the standard procedure
is to use perturbation theory from which one can compute the lowest order interaction correction coefficient, $\mathcal{C}$, to the conductivity:
\be
\sigma = \sigma_0 \Big ( 1 + \alpha\, \mathcal{C} + \Ord \left( {\bf \alpha}^2\right) \Bigr)\, ,
\ee
which is related to the two-loop correction to the polarization operator, see the corresponding diagrams in Fig.~\ref{fig:rqed-2loop-polarization}.
The later is defined as:
\begin{subequations}
\label{Pi2-def}
\bea
&&\Pi_2^{\mu \nu}(q) = 2\,\Pi_{2a}^{\mu \nu}(q) + \Pi_{2b}^{\mu \nu}(q)\, ,
\label{pi-2l} \\
&&i\Pi_{2a}^{\mu \nu}(q) =
- \int [\D^{d_e}k]\, \Tr \left[ (-ie\gamma^\nu) S_0(k+q)(-ie\gamma^\mu)S_0(k)\left( -i \Sigma_{1}(k) \right) S_0(k) \right]\, ,
\label{pi-2la}\\
&&i\Pi_{2b}^{\mu \nu}(q) =
\label{pi-2lb} - \int [\D^{d_e}k_2]\,\Tr \left[ (-ie\gamma^\nu) S_0(k_2+q)(-ie\Lambda_1^\mu(k_2,q))S_0(k_2) \right]\, ,
\nonum
\eea
\end{subequations}
where $\Sigma_1(k)$ is the one-loop self-energy, Fig.~\ref{fig:one-loop}b, and $\Lambda_1^\mu(k_2,q)$ is the one-loop vertex part, 
Fig.~\ref{fig:one-loop}c, which are defined as:
\begin{subequations}
\bea
&&-i\Sigma_1(k) = \int [\D^{d_e}p]\, (-ie\gamma_\mu) S_0(k+p)(-ie\gamma_\nu) \tilde{D}_0^{\mu \nu}(p)\,,
\label{sigma-1l}
\\
&&-ie\Lambda_1^\mu(k_2,q) = \int [\D^{d_e}k_1]\, \tilde{D}_0^{\al \beta}(k_1-k_2)\, (-ie\gamma_\al)S_0(k_1+q)(-ie\gamma^\mu)S_0(k_1)(-ie\gamma_\beta)\, .
\label{vertex-1l}
\eea
\end{subequations}
%
We now review the computations of these diagrams and the corresponding optical conductivity in the non-relativistic and relativistic limits.

\FloatBarrier


In the non-relativistic limit ($v \ll c$), most results obtained by computing these diagrams seem to agree with the
result first found by Mishchenko~\cite{Mishchenko08}:
\be
\mathcal{C} = \frac{19-6\pi}{12} \approx 0.013 \qquad (v \ll c)\, .
\label{C-NR}
\ee
Interestingly, this result shows that $\al_g \mathcal{C} \approx 0.02$ in agreement with the small deviations seen experimentally. 
Moreover, from Eq.~(\ref{sigma00}), Mishchenko found that this result is decomposed as: 
$\mathcal{C} = \mathcal{C}_a + \mathcal{C}_b$ where $\mathcal{C}_a=1/4$ is the contribution of diagrams  Fig.~\ref{fig:rqed-2loop-polarization}a
and $\mathcal{C}_b=(8-3\pi)/6$ is the contribution of the diagram  Fig.~\ref{fig:rqed-2loop-polarization}b. 
We therefore notice that the ``complicated'' contribution, that is the one proportional to $\pi$
in Eq.~(\ref{C-NR}), arises from  the ``complicated'' diagram Fig.~\ref{fig:rqed-2loop-polarization}b.


Less relevant to experiments but still of academic interest is to compute the interaction correction coefficient at the infra-red fixed point. 
Performing the trace algebra in Eqs.~(\ref{Pi2-def}) and after Wick rotation to go to euclidean space ($q_0 = \I q_{E0}$) the diagrams reduce to a combination of scalar integrals of the kind
encountered in the previous sections. All of these integrals reduce to simple loops and chains except for one, arising from diagram b in Fig.~\ref{fig:rqed-2loop-polarization}, and which
is precisely given by $I(\lambda)$ defined in (\ref{def:I}).
All calculations done, the two-loop correction to the polarization operator reads (for $\veps_\gamma \ra 0$):
\begin{subequations}
\bea
\Pi_{2}(q^2) =&& 4 N_F\,
\frac{e^4 \, \Gamma(\lambda)}{(4\pi)^{3+\lambda} \, \left(-q^2\right)^{1-\lambda}}
\frac{16 \Gamma(1+\lambda) \Gamma(1-\lambda)}{ \Gamma(3+2\lambda)} \, C_1(\lambda) \, ,
\label{2loop-a+b} \\
C_1(\lambda) =&& 2\lambda - \frac{5}{2} - \frac{3}{2\lambda} + \frac{1}{1+\lambda}
+ \frac{3}{2} \lambda (1+\lambda) \Bigl[ \psi'(\lambda) - \psi'(1) \Bigr] \, .
\label{C1}
\eea
\end{subequations}
where the result of Eq.~(\ref{result:I}) has been used. 
Adding the one-loop contribution yields:
\begin{subequations}
\bea
\label{1+2loopdiagram}
\Pi(q^2) &&= \Pi_{1}(q^2) \Bigl(1+\alpha C(\lambda) + \Ord \left( {\bf \alpha}^2\right)
\Bigr),  \\
~~ C(\lambda) &&= - \frac{1}{\pi\lambda(1+\lambda)} \, C_1(\lambda)
= - \frac{1}{2\pi} \left(3 \Bigl[ \psi'(\lambda+2) - \psi'(1) \Bigr]
+ \frac{4}{1+\lambda} + \frac{1}{(1+\lambda)^2} \right)
 \, .
\eea
\end{subequations}
In the case $\lambda=1/2$, this leads to:~\cite{KotikovT13,Teber12}
\be
\mathcal{C}^* = C(1/2) = \frac{92-9\pi^2}{18\pi} \approx 0.056 \qquad (v = c)\, .
\label{C-R}
\ee
So in the relativistic limit, the overall correction to the conductivity is even smaller than in the non-relativistic limit as: $\al \mathcal{C}^* \approx 4.10^{-4}$.
Interestingly, the ``complicated'' contribution in Eq.~(\ref{C-R}) is precisely the same as in Eq.~(\ref{C-NR}). The difference between these two results
seems to arise essentially from the simple diagrams of Fig.~\ref{fig:rqed-2loop-polarization}a. The latter contain the sub-divergent one-loop fermion self-energy which is 
responsible for the flow of the Fermi velocity. 

\begin{figure}
    \includegraphics[width=0.9\textwidth]{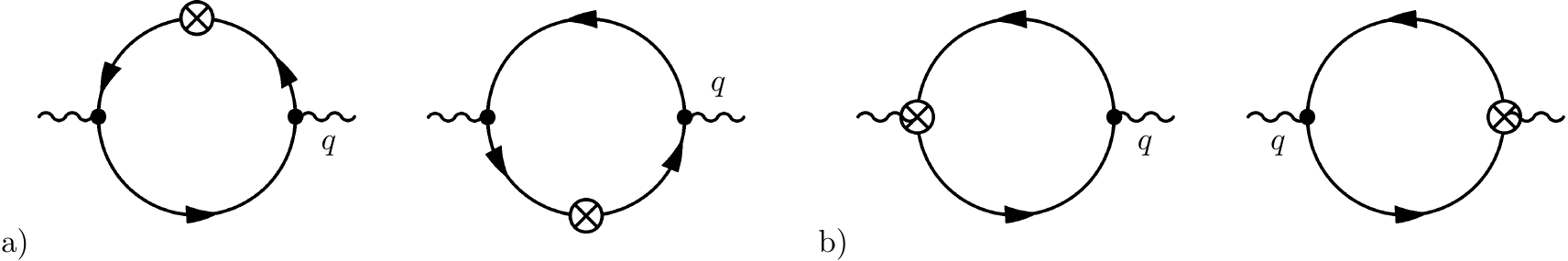}
    \caption{\label{fig:2loop-polarization-ct}
     Counter-term diagrams to the two-loop vacuum polarization diagrams.}
\end{figure}

In the case $v \ll  c$, the result of Eq.~(\ref{C-NR}) contains contributions from the counter-terms of the diagrams in
Fig.~\ref{fig:rqed-2loop-polarization} (see discussions in Ref.~[\onlinecite{C-theory,TK14}]).
We will now show that such contributions are absent in the relativistic limit $v=c$.
The counter-term diagrams are represented graphically in Fig.~\ref{fig:2loop-polarization-ct} and include four contributions:
\be
\tilde{\Pi}_2^{\mu \nu}(q) = 2\tilde{\Pi}_{2a}^{\mu \nu}(q) + 2\tilde{\Pi}_{2b}^{\mu \nu}(q) \, .
\label{counter}
\ee
The first term, $2\tilde{\Pi}_{2a}$, comes  from the singular part of the internal loops
in the two first diagrams in Fig.~\ref{fig:rqed-2loop-polarization}a, that is the one-loop fermion self-energy of Fig.~\ref{fig:one-loop}b.
The second term, $2\tilde{\Pi}_{2b}$, comes from the singular part of two internal triangles in the last diagram, that is
the one-loop vertex part of Fig.~\ref{fig:one-loop}c. These one-loop graphs were computed in Ref.~[\onlinecite{Teber12}] and we reproduce the results here for clarity:
\begin{subequations}
\label{1l-sig+vert}
\bea
&&\Sigma_{1V}(k^2) = -\frac{e^2 \,\Gamma(\lambda) (-k^2)^{-\varepsilon_\gamma}}{(4 \pi)^{d_\gamma/2}}
\left[ \frac{2(\lambda-\varepsilon_\gamma)^2}{1 + \lambda-2\varepsilon_\gamma} - \xi\,(\lambda-\varepsilon_\gamma) \right]\,
A(1,\lambda)\, ,
\label{sigma-1l-1}
\\
&&\Lambda_1^\mu = \frac{e^2 \,\Gamma(\lambda) \, m^{-2\varepsilon_\gamma}}{(4\pi)^{d_\gamma/2}}\,\gamma^\mu\,
\left[ \frac{2(\lambda-\varepsilon_\gamma)^2}{1 + \lambda-\varepsilon_\gamma } - \xi \lambda \,\right]\,
\frac{\Gamma(\varepsilon_\gamma)}{\Gamma(1+\lambda)} \, ,
\label{vertex-1l-1}
\eea
\end{subequations}
where we have used the parametrization $\Sigma_1(k) = \Sk \Sigma_{1V}(k^2)$, $A(\al,\beta)$ was defined in Eq.~(\ref{def:A})
and the mass $m$ has been used as an infra-red regulator in the expression of the vertex correction, Eq.~(\ref{vertex-1l-1}).

The results for $\tilde{\Pi}_{2a}$ and $\tilde{\Pi}_{2b}$ can then be represented as:
\be
\tilde{\Pi}_{2a}^{\mu \nu}(q) = -\text{Sing}\,\left[\Sigma_{1V}(k^2)\right]\, \Pi_{1}^{\mu \nu}(q) \, ,
\qquad
\tilde{\Pi}_{2b}^{\mu \nu}(q) = -\text{Sing}\,\left[\Lambda^{\mu}/\gamma^{\mu}\right] \Pi_{1}^{\mu \nu}(q) \, ,
\label{Pi2a+b-ct}
\ee
where $k$ is the internal momentum, $\Pi_{1}^{\mu \nu}(q)$ is the one-loop contribution, Eq.~(\ref{1loop}) 
and the renormalization of the one-loop vertex has been represented as: $\text{Sing}\,\left[\Lambda^{\mu}\right] \equiv \gamma^{\mu}\, \text{Sing}\,\left[\Lambda^{\mu}/\gamma^{\mu}\right]$.
In the improved minimal subtraction scheme ($\overline{\rm{MS}}$), for any function $F(\epsilon_{\gamma})$, the Sing operator is defined as:
\be
\text{Sing}~\left[F(\epsilon_{\gamma})\right] \equiv \frac{1}{\epsilon_{\gamma}} F(\epsilon_{\gamma}=0),~~~
\epsilon_{\gamma}=(4-d_{\gamma})/2 \, .
\label{Sing}
\ee
With the help of this definition and from Eqs.~(\ref{1l-sig+vert}) we obtain:
\begin{subequations}
\bea
\text{Sing}\,\left[\Sigma_{1V}(k^2)\right] &=& -\frac{e^2}{(4\pi)^2} \left[\frac{2\lambda}{1+\lambda}-\xi\right] 
\frac{1}{\epsilon_{\gamma}} \, ,
\label{SingSigma}
\\
\text{Sing}\,\left[\Lambda^{\mu}/\gamma^{\mu}\right] &=& +\frac{e^2}{(4\pi)^2} \left[\frac{2\lambda}{1+\lambda}-\xi\right]
\frac{1}{\epsilon_{\gamma}} \, .
\label{SingVertex}
\eea
\end{subequations}
Eqs.~(\ref{SingSigma}) and (\ref{SingVertex}) are exactly opposite to each other. The contribution of the counter-terms, Eq.~(\ref{counter}),
is therefore zero. This actually follows from the Ward identity for reduced QED which relates the wave function and vertex renormalization constants.

\acknowledgments

One of us (A.V.K.) was supported by RFBR grant 16-02-00790-a. 
The authors thank the Organizing Committee of V International Conference ``Models in
Quantum Field Theory'' (MQFT-2015) for their invitation.


\end{fmffile}
\end{document}